# Pairing, pair-breaking, and their roles in setting the $T_c$ of cuprate high temperature superconductors

T. J. Reber[1], S. Parham[1], N. C. Plumb[1,†], Y. Cao[1], H. Li[1], Z. Sun[1,&], Q. Wang[1], H. Iwasawa[2], M. Arita[2], J. S. Wen[3], Z. J. Xu[3], G.D. Gu[3], Y. Yoshida[4], H. Eisaki[4], G.B. Arnold[1], D. S. Dessau[1]*

[1] Dept. of Physics, RASEI, and JILA, University of Colorado, Boulder, 80309-0390, USA.

[2] Hiroshima Synchrotron Radiation Center, Hiroshima University, Higashi-Hiroshima, Hiroshima 739-0046, Japan.

[3] Condensed Matter Physics and Materials Science Department, Brookhaven National Labs, Upton, New York, 11973 USA.

[4] AIST Tsukuba Central 2, 1-1-1 Umezono, Tsukuba, Ibaraki 3058568, Japan.

† Now at Swiss Light Source, Paul Scherrer Institut, CH-5232 Villigen PSI, Switzerland.

& Now at National Synchrotron Radiation Center, Hefei, China

* Correspondence to: dessau@colorado.edu

**The key ingredients in any superconductor are the Cooper pairs, in which two electrons combine to form a composite boson. In all conventional superconductors the pairing strength alone sets the majority of the physical properties including the superconducting transition temperature $T_C$. In the cuprate high temperature superconductors, no such link has yet been found between the pairing interactions and $T_C$. Using a new variant of photoelectron spectroscopy we measure both the pair-forming ($\Delta$) and a self energy/pair-breaking term ($\Gamma_S$) as a function of sample type and sample temperature, and we make the measurements over a wide range of doping and temperatures within and outside of the pseudogap/competing order doping regimes. In all cases we find that $T_C$ is approximately set by a crossover between the pair-forming strength $\Delta$ and 3 times the self-energy term $\Gamma_s$. – a new paradigm for superconductivity. In addition to departing from conventional superconductivity in which the pairing alone sets $T_C$, these results indicate the zero-order importance of the near-nodal self-energy effects compared to competing order/pseudogap effects.**

The macroscopic quantum state that defines a superconductor is a condensate of Cooper pairs, which are two electrons bound together to form a composite boson. With increasing temperature, thermal fluctuations get stronger and stronger, eventually overcoming the binding energy holding the pairs together (the pairing energy $2\Delta$). This breakup of the pairs typically sets the superconducting transition temperature $T_C$, with the $T_C$ directly related to the strength of the pairing $2\Delta$. Concomitant with these ideas is a direct relation between the pairing scale and Tc, i.e. $2\Delta/kT_C \sim 3.52$ for the weak-coupling Bardeen-Cooper-Schreiffer case, with this ratio increasing to 5 or more for so-called "strong coupling" superconductors[1]. As it sets the $T_C$ of the superconductor, this pairing scale $\Delta$ effectively becomes the only relevant energy scale for the problem.

In the high temperature cuprate superconductors the pairing energy is much stronger than in conventional superconductors and so it is natural that these pairs survive to much higher temperatures. Surprisingly however, it has been found that for many of the samples (especially those in the underdoped region of the phase diagram) these pairs exist even above the transition temperature (to a temperature $T_{pair}$)[2,3,4,5]. And while there is great interest in understanding the regime between $T_{pair}$ and $T_C$ as well as what sets the phase coherence temperature, little direct information exists to date.

A great intensity of work over the past two decades has focused on the pseudogap and the T* line of the doping phase diagram, as this was thought by many to represent a prepairing state and a precursor to the SC. Along with this, the larger gaps in the UD side suggested stronger SC pairing, though recent results indicate that these larger gaps are likely not directly related to the SC at all [6,7]. Recently, a great amount of attention has focused on CDW's in the underdoped regime, which would most likely be a competing phase of the superconductivity [8,9]. With all of this intense work, there surprisingly has been little work looking for similarities between the under- and over-doped regimes, perhaps because it is generally thought that these two regimes are so dramatically different. Here we show that in fact critical aspects of the superconducting physics are the same for the UD and OD samples especially if we consider the electronic self-energy or scattering rate effects.

Here, using a new variant of angle-resolved photoemission (ARPES) over a wide range of doping levels, we show that there is an additional key parameter that helps control the $T_C$ in these materials. We use the new TDoS (Tomographic Density of States) technique, which has previously been shown to have the ability to quantitatively extract both the pairing and pair-breaking energy scales as a function of temperature [4,5]. A major advantage of the TDOS method is that it allows us to effectively remove the major impact of the heterogeneity on the ARPES spectra, giving us what is essentially the homogenous self-energy term $\Gamma_s$ [4], which is significantly smaller than the actual linewidth of the individual ARPES spectra as these are seriously affected by the doping or chemical-potential heterogeneity (also see supplemental materials). By doing this procedure on a wide range of sample types from underdoped to overdoped and including both single and double-layer samples with drastically different $T_C$'s, we empirically show that $T_C$ is determined by the crossover with temperature of these two energy scales. This result, which is very different from the BCS picture, connects much of the known phenomenology of the cuprates.

Figure 1 shows raw TDoS data as a function of temperature from 8 different sample types, including uderdoped (UD), optimally doped (Opt) and overdoped (OD) bilayer $Bi_2Sr_2CaCu_2O_8$ samples with Tc values listed as the prefix of the sample name (OD75K is an overdoped sample with Tc=75K). Also included is a single-layer $Bi_2Sr_2CuO_6$ sample with Tc=28K and a Ni-impurity doped $Bi_2Sr_2CaCu_2O_8$ sample with Tc=80K. All data were taken along a cut perpendicular to the Fermi surface that is 12 degrees away from the node (N), where the gap is zero (inset of panel e). As described elsewhere, the one-dimensional integration inherent in the TDoS analysis turns the data into k-localized density of states curves, which may then be fit by the following form first used by Dynes to fit tunneling spectra of conventional superconductors [10].

$$I_{TDoS} = \text{Re} \frac{\omega - i\Gamma_s}{\sqrt{(\omega - i\Gamma_s)^2 - \Delta^2}} \qquad (1)$$

This is essentially just a BCS density of states with pairing gap $\Delta$ broadened by the term $\Gamma_s$ which can be equivalently looked at as the pair-breaking scattering rate or as the single-particle electron self-energy term as described within the Nambu-Gorkov formulation of superconductivity (see supplemental materials). Fig S1 in the Supplemental Materials shows a direct comparison of the fits to all curves shown in Fig 1. Using only the two parameters of equation 1 ($\Delta$ and $\Gamma$s), the fit quality is excellent in all cases.

While Fig 1 shows TDoS data taken at one Fermi surface angle (12 degrees), we in fact utilized and fit data from many angles for each sample. Fig S2 in the supplemental materials shows an example full set of data used on one sample, containing 6 cuts at different Fermi surface angles $\theta_{FS}$ multiplied by 10 different temperatures, giving ~ 300,000 data points total per sample. These are reduced to the specific plots of $\Delta$ and $\Gamma$s vs. k for each temperature (right plots of fig S2) or vs. temperature for each k (bottom plots of fig S2), confirming the d-wave nature of $\Delta$ and the near-nodal approximate k-independence of $\Gamma_s$ for all temperatures. With this experimental knowledge of the rate at which $\Delta$ grows with Fermi surface angle $\theta_{FS}$ (often termed $v_\Delta$) we can use a simple d-wave form to extrapolate to the superconducting gap maximum $\Delta_{Max}$ that in principle would occur at the antinode in the absence of any pseudogapping effects.

$$\Delta(\theta_{FS}) = \Delta_{Max} |\sin(2\theta_{FS})| \tag{2}$$

This value of $\Delta_{Max}$ may be different than the gap directly measured at the antinode, and we argue that this near-nodal or $v_\Delta$ gap is a more accurate way to determine the superconductive pairing scales than the direct measurement of $\Delta$ at the antinode, with the latter contaminated by the antinodal pseudogap [6,7]. The net result of doing this at each temperature is the plot of $\Delta_{Max}$ and $\Gamma_s$ versus T, as shown in the lower right panel of fig S2. The great many experimental data points per temperature on this sample have thus been reduced to the two key parameters ($\Delta_{Max}$ and $\Gamma_s$) per temperature in this plot. A qualitatively similar temperature dependent growth has previously been extracted for optimally doped samples from STM measurements (11), our TDoS ARPES [5], and other very recent ARPES [12].

Figure 2 shows a similar compilation of the data from all the fits on all the samples, each carried out in an identical way as for the optimally doped sample shown in Figs S1 and S2. In general, the magnitude of $\Gamma_s$ found here is huge compared to the magnitude of $\Gamma_s$ seen in other strongly coupled superconductors, either in raw numbers or when rescaled to the gap values. The left axes ($\Delta_{Max}$) of fig 2 have been scaled by a factor of three compared to the right axes ($\Gamma_s$). With this done it is seen that $\Delta_{Max}$ and $3\Gamma_s$ cross very near $T_C$ for every sample, including under-doped, optimally-doped, over-doped, impurity-doped and both single and double-layer cuprates, with $T_C$'s ranging from 28K to 91K. That this empirical finding holds over such a diversity of materials indicates that it is extremely unlikely to be a coincidence, and is likely the main ingredient for what lowers the $T_C$ of the cuprates from the onset of pairing $T_{Pair}$. Together with $\Delta_{Max}$ it is apparently what sets the $T_C$ of the cuprate superconductors.

Figure 3 contrasts the behavior of the cuprates uncovered here with that of a standard BCS superconductor, with the main difference being the filling in of the gap due to the pair-breaking interactions, rather than the simple BCS-like gap closing that occurs in the absence of the temperature-dependent pair-breaking. The filling in of the gap with increasing temperature is also observed in optics (13), tunneling (14), and thermodynamics (15), and thus appears to be <u>a generic feature of the cuprates</u> that is fully consistent with the increase of $\Gamma_s$ with temperature that we detail here. Further, the disappearance of the "coherence peak" near and above Tc observed in earlier ARPES (16,17) and tunneling (14) experiments is also fully consistent with the rapid increase of $\Gamma_s$ with temperature, though none of these earlier measurements were able to quantify the strength of this effect. Essentially all other main spectroscopic features observed in the cuprates are also fully consistent with our data and the schematic of figure 3b, including the findings that a) the gap energy scales stay approximately constant below $T_C$ (4, 13,14,16), b) there are prepairing fluctuations persisting for up to 30K above $T_C$ (2,5), and c) there are "Fermi arcs" in underdoped samples (18) with length increasing with temperature (19), as explained in (4). Therefore, the present result explains the general behavior of these many spectroscopies, with the new quantitative results allowing us to obtain a much greater in-depth understanding.

We can intuitively understand how the competition of pairing and pair breaking sets the Tc by imagining that the pair-breaking processes at high temperatures reduce the pair lifetimes to the point that the pairs are broken almost as rapidly as they are formed.[20]

Therefore, the overall pair density just below $T_{Pair}$ is relatively low, even though the pair binding strength (the gap energy $\Delta$) is still rather large. As we lower the temperature towards $T_C$ the pairing strength increases slightly, but more importantly, the lifetimes of the pairs increases as seen directly by the decrease in the pair fluctuation strength $\Gamma_s$. At approximately the ratio $\Delta/\Gamma_s=3$, the pairs are dense enough and long enough lived that phase coherence between them may occur. Ultimately it is this onset of long-range phase coherence that marks the onset of the critical temperature $T_C$ – that this is set approximately by the ratio $\Delta(T)/\Gamma_s(T)=3$ for all doping levels is one of the major new results of this work.

A dependence of Tc on the ratio of pairing and pair-breaking has been well described theoretically[21], particularly in cases of induced pair-breaking through the introduction of disorder or magnetism. For example, similar physics has in fact been observed in the context of granular aluminum very near the metal-insulator transition, where localization and enhanced interaction effects become critically important, and $\Gamma_s$ can be continuously tuned by changing the size of the aluminum grains. Tunneling measurements, analyzed using formula 1 showed that superconductivity disappears when the $\Gamma_s$ term becomes comparable to $\Delta$([22]). Of course, the pair-breaking $\Gamma_s$ in granular Al is a static effect, while the dynamic (temperature dependent) $\Gamma_s$ seen in the cuprates is a more unusual and important case, and the ratio of approximately 3 in the present materials is not at the moment understood.

There are two main classes of explanations for a strongly temperature-dependent $\Gamma_s$. A) $\Gamma_s$ represents the large self-energy effects that are present in the normal state of the cuprates, with these being largely gapped out in the superconducting state, i.e. the superconducting state "undresses" these correlations[23,24]. In this picture these correlations are recovered as the temperature is raised towards the normal state, with these correlations being responsible for the $\Gamma_s$ term and the filling in of the gap. B) $\Gamma_s$ represents thermally excited excitations out of the superconducting state, for example quasiparticles across the d-wave gap edge, phonons, magnons or the magnetic resonance mode [25,26], or topological vortex excitations similar in spirit to what occurs in the Kosterlitz-Thouless transition. While more work is needed to conclusively determine the origin of the strong pair-breaking, the fact remains that the strongly temperature-dependent pair-breaking is a critical aspect of the physics, which appears to help set the $T_C$ of the cuprates. Viewed from a different angle, finding a way to minimize or even remove these pair-breaking interactions points out a route towards increasing the transition temperatures of these materials, with an available temperature window for increasing the $T_C$ that is upwards of 20K.

**Acknowledgments:** We thank A. Balatsky, E. Calleja, A. Chubukov, M. Hermele, K. McElroy, L. Radzihovsky and Xiaoqing Zhou for valuable conversations and D. H. Lu and R. G. Moore for help at SSRL. SSRL is operated by the DOE, Office of Basic Energy Sciences. ARPES experiments at the Hiroshima Synchrotron Radiation Center were performed under proposal 09-A-48. Funding for this research was provided by DOE Grant No. DE-FG02-03ER46066 (Colorado) and DE-AC02-98CH10886 (Brookhaven).

**Author Contributions:** T.J.R. And D.S.D. designed the experiment. J.S.W., Z.J.X, G.G., Y.Y., and H.E. grew and characterized the single crystal samples. T.J.R., S.P., H.L., N.C.P., Y.C., Q. W., M.A. and H.I. acquired the experimental data. T.J.R., S.P., H.L. and N.C.P. performed the data analysis. G.B.A. provided theoretical support. T.J.R. and D.S.D wrote the manuscript. All authors discussed the results and commented on the manuscript.

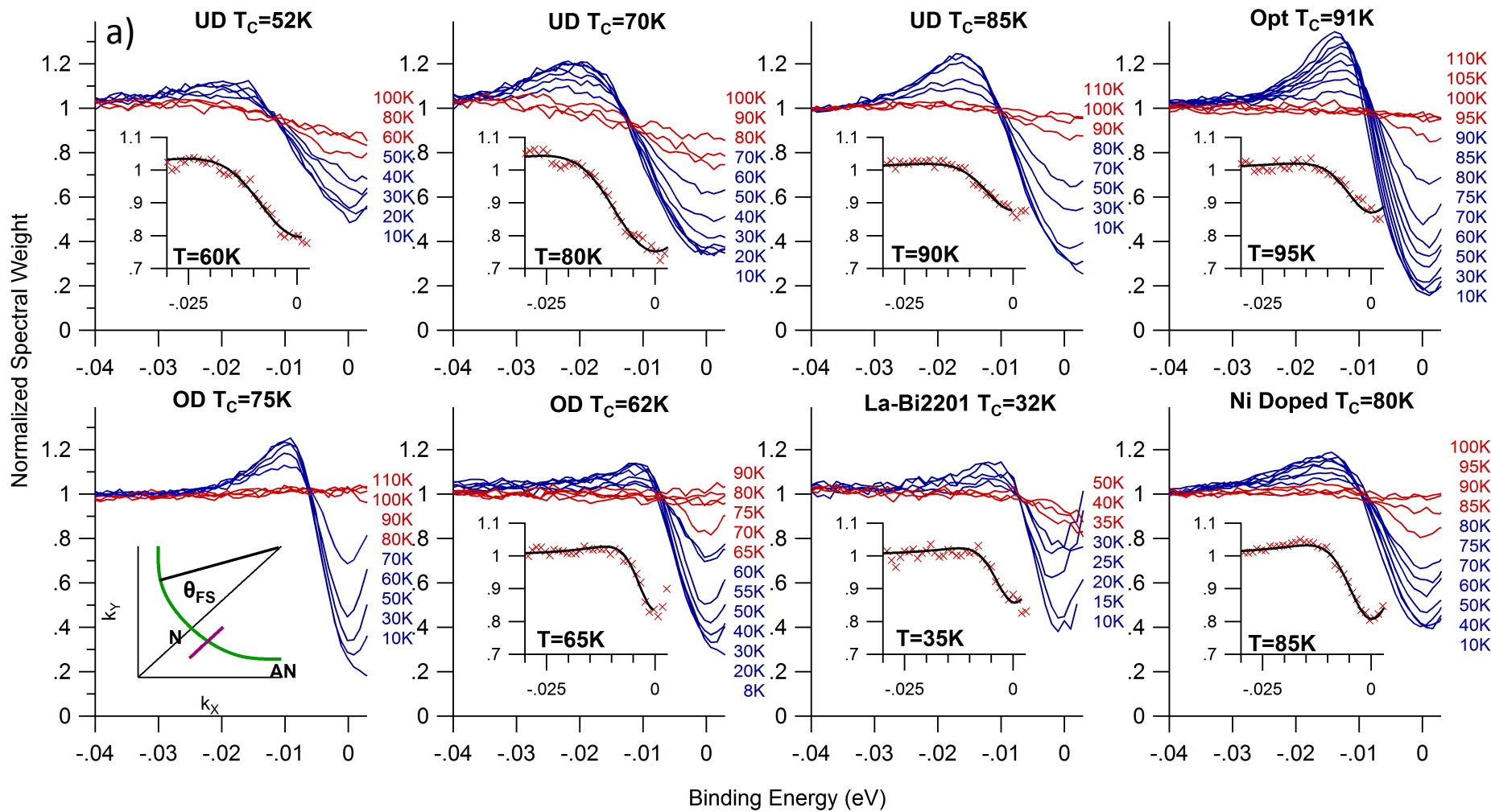

**Fig 1.** Temperature dependent TDoS ARPES data of $Bi_2Sr_2CaCu_2O_{8+d}$ (Bi2212) along a cut 12 degrees away from the node N (inset of lower left panel shows the cut location in the Brillouin zone). Data from 8 samples are shown, labeled with their $T_C$ values and doping types (UD80K = an underdoped Bi2212 sample with Tc=80K), Opt=optimally doped, OD=overdoped. All samples are double-layer Bi2212 except for panel g which is single-layer $Bi_2Sr_2CuO_6$ (Bi2201). Blue curves are for T<Tc while red curves are for T>Tc, with many of these T>Tc curves still showing a weak gapping effect. The insets show expanded views of data for T slightly greater than Tc, including fits to those data using equation 1 (black lines through the curves).

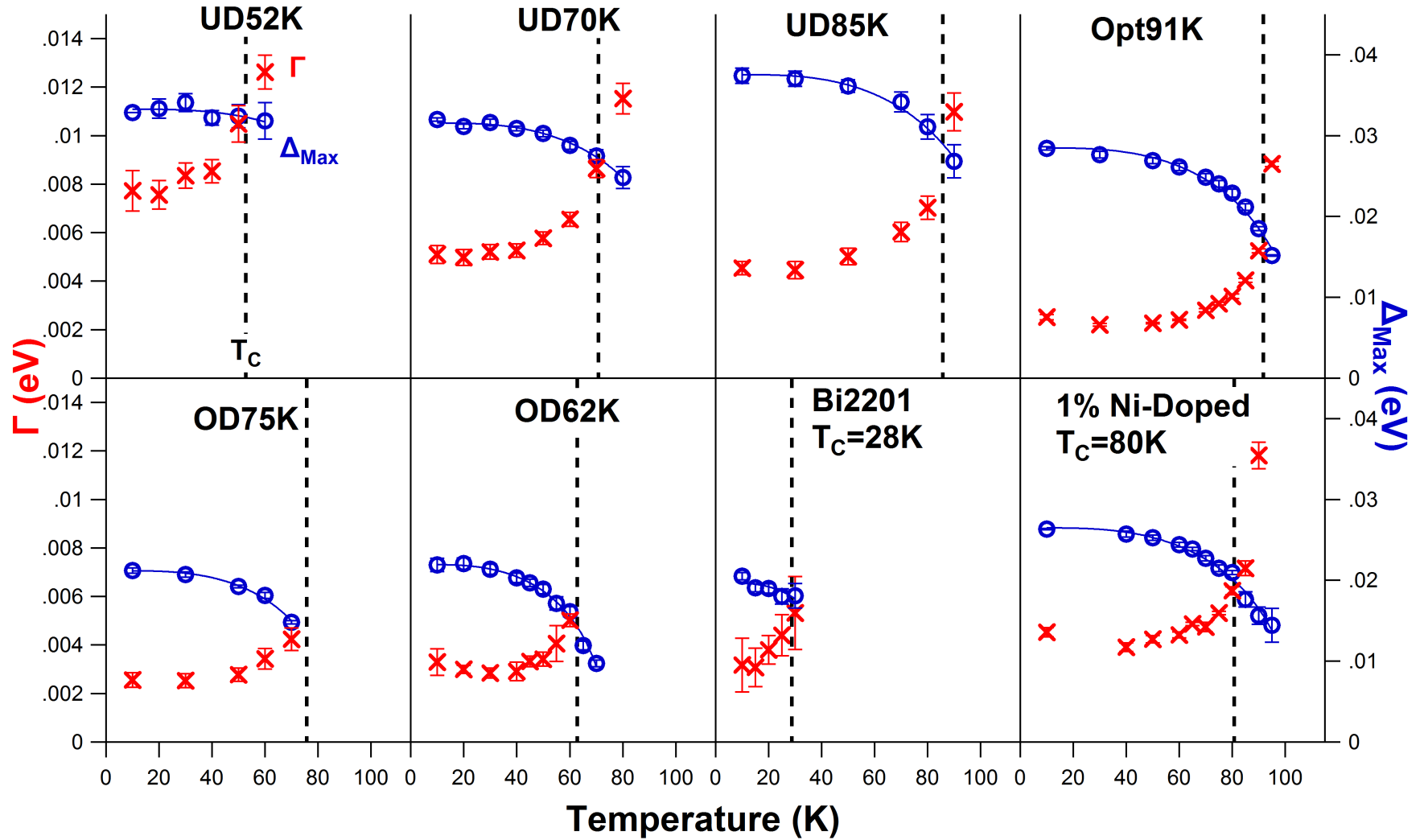

**Fig 2.** A compilation of $\Delta_{Max}$ (left axis) and $\Gamma$ (right axis) for the 8 samples shown in figure 1, from fits to TDoS data over a full grid of many temperatures and angles (see for example Fig S1 for a grid on one sample). $\Delta_{Max}$ is the d-wave maximum superconducting gap determined from the near-nodal data (e.g. $v_\Delta$) while $\Gamma$ is the average zero-frequency pair-breaking self energy over the near-nodal region. Error bars are shown in all cases though are difficult to see when they are smaller than the symbol size.

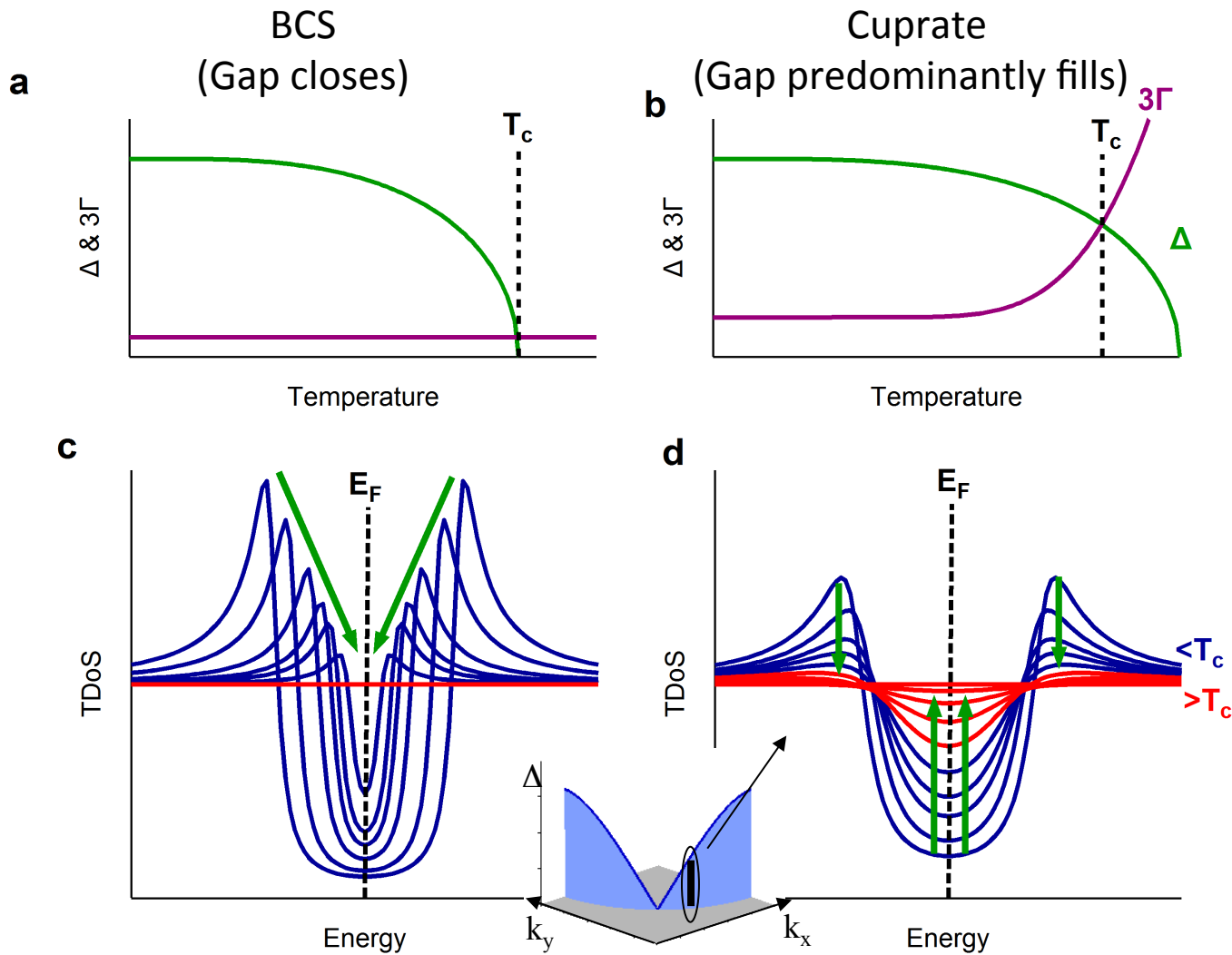

**Fig 3.** Contrast between the temperature evolution of a BCS superconductor (left) and a cuprate (right). Blue=superconducting state (T<$T_C$) while red= normal or pseudogapped state (T>$T_C$). (a,b) With increasing temperature the BCS gap closes, reaching zero at T=$T_C$. The pair-breaking scattering rate $\Gamma_S$ is in general very small and almost temperature-independent, so it is almost always considered unimportant. (d) With increasing temperature the gap in cuprates principaly fills, which is due to the rapidly rising $\Gamma_S$ with temperature (panel c). The gap filling is a phenomenology observed in essentially all spectroscopies on cuprates, but has been difficult to quantify until now. Here, the superconducting transition is not marked by the gap closing to zero, but rather by a significant filling, roughly determined by the ratio $\Delta_{Max}/3\Gamma_S \sim 1$.